
\magnification=\magstep1  \overfullrule=0pt
\def\la{\lambda} \def\om{\omega} \def\sig{\sigma} 
\def\al{\alpha}  \def\Eqde{\,\,{\buildrel {\rm def} \over =}\,\,}
\def\be{\beta}    \def\r{\bar{r}}    \def\A{A^{(1)}}
\font\huge=cmr10 scaled \magstep2

{\nopagenumbers
\rightline{August, 1994}
\bigskip\bigskip
\centerline{{\bf \huge The Classification of $\widehat{SU}(m)_{k}$}}
\bigskip \centerline{{\bf \huge Automorphism Invariants}}\bigskip
\bigskip\bigskip
\centerline{Terry Gannon}
\bigskip\centerline{{\it Institut des Hautes Etudes Scientifiques}}
\centerline{{\it 91440 Bures-sur-Yvette, France}}\bigskip\bigskip
\noindent{{\bf Abstract}}. In this paper we find all permutations of the
level $k$ weights of the affine algebra $\A_r$ which commute with both its
$S$ and $T$ modular matrices. We find that all of these are simple
current automorphisms and their conjugations. Previously, the $\A_{r,k}$
automorphism invariants were known only for $r=1,2$ $\forall k$,
and $k=1$ $\forall r$. This is a major step toward the full classification
of all $\A_{r,k}$ modular invariants; the simplicity of this proof strongly
suggests that the full classification should
be accomplishable. In an appendix we collect some new results concerning
the $\A_{r,k}$ fusion ring.
\vfill \eject} \pageno=1

There has been much speculation in recent years about the ``coincidences''
appearing in the classification of modular invariants in WZW conformal
field theory, from A-D-E [1], to Fermat curves [2], to generalized
Coxeter graphs [3].
For these reasons, as well of course for the classification of CFT itself,
it is certainly of interest to find all possible partition functions
for WZW theories, based on the affine algebra $\A_r$ at level $k$.

In particular, the problem is to find all modular invariant
sesquilinear combinations of $\A_{r,k}$ characters
$$Z=\sum_{\la,\mu}M_{\la\mu}\,\chi_\la\,\chi_\mu^*,\eqno(1)$$
where each $M_{\la\nu}$ is a non-negative integer, and where the sum
is over all $\rho$-shifted weights $P_{++}(\A_{r,k})$ ($\rho=
{1\over 2}\sum_{\al>0}\al$). In addition there is the condition that
$M_{\rho\rho}=1$. Any such matrix $M$ will be called a {\it physical
invariant}.

We probably already know all but finitely many physical invariants for
$\A_{r,k}$ [4-8]. Almost all of them are {\it simple current invariants} [4-6]
and their conjugations, which we can call ${\cal D}$-type invariants.
These can be counted [9] -- e.g.\ for any odd
prime $p$ we expect exactly 4 physical invariants for $\A_{p-1,k}$ at most
$k$. The remaining physical invariants are called {\it exceptional}, and
can be collected into two classes: those corresponding to exceptional
chiral extensions (called ${\cal E}_{6}$-type exceptionals); and those
corresponding to exceptional automorphisms of simple current extensions
(called ${\cal E}_{7}$-type exceptionals).
There will be $\A_{r,k}$ exceptionals at levels $k=r-1$ (if $r\ge 4$), $r+1$
(if $r\ge 3$), and $r+3$ (if $r\ge 2$), all due to
conformal embeddings [7]. Apart from these, there
are only finitely many remaining known exceptionals [7,8].

The most famous classification
is the A-D-E one for $\A_{1,k}$ [1,10]. There we find an ${\cal A}_k$-series
(for all $k$) and ${\cal D}_k$-series (for even $k$) -- these are
of ${\cal D}$-type. There are also three exceptionals, at levels 10, 28 (both
are of ${\cal E}_{6}$-type) and 16 (${\cal E}_{7}$-type).
The only other classifications at present for $\A_{r,k}$ physical invariants
are at $k=1$ for all $r$ [11], and $r=2$ for all $k$ [12].

The general problem reduces naturally to two parts:

\item{(i)} find all possible chiral extensions of the $\A_{r,k}$ current
algebra (this corresponds essentially to finding all possibilities for
the values $M_{\rho\la},M_{\la\rho}$);

\item{(ii)} find all automorphism invariants for each of these extensions.

In this paper we address part (ii). We will focus exclusively on the
unextended chiral algebra, but it can be expected that this argument
will generalize to the automorphism invariants of all simple current
extensions of $\A_{r,k}$. Indeed, one of the two tasks suggested by this
paper is to explicitly find all simple current extensions and their
automorphisms.

We will prove that the only automorphism invariants for $\A_{r,k}$ are
the ``obvious'' ones. Incidently, this is not the case for all affine
algebras: e.g.\ there are exceptional automorphism invariants for
$G_{2,4}^{(1)}$ and $F_{4,3}^{(1)}$ [8]. The other natural task suggested
by this paper is to find all automorphism
invariants for the other simple affine algebras [13]. The methods developed
here should also apply to them, with little change.

We will find that there are precisely
$${\cal NA}\,(\A_{r,k})=2^{c+p+t}\eqno(2)$$
automorphism invariants for $\A_{r,k}$, where:

\item{$c$}$=1$, unless $r=1$ or $k\le 2$ when $c=0$ (an exception: $c=-1$ when
both $r=1$ and $k=2$);

\item{$p$} is the number of distinct odd primes which divide $r+1$
but not $k$;

\item{$t$}$=0$ if either $r$ is even, or $r$ is odd and $k\equiv 0$
(mod 4), or $k$ is odd and $r\equiv 1$ (mod 4) -- otherwise $t=1$.

Other large classes of automorphism invariants are classified in
[14] for $\A_{1,k_1}\oplus\cdots\oplus\A_{1,k_r}$, and in [15] for ``simple
current automorphism invariants'' for any RCFT (subject to a condition on
the corresponding $S$ matrix). The argument given here is
completely novel; in particular the argument in [15] does not apply here
because it assumes the automorphism invariant is ``locally'' given
by a simple current (not true for $\A_{r,k}$), and it also assumes
that the $S$ matrix does not have too many zeros -- something which
is still unproven for $\A_{r,k}$ it seems.

The simplicity of the proof in this paper puts into question the pessimism
surrounding the possibility of finding a general classification of all
physical invariants for
$\A_{r,k}$ -- or indeed for $g_k$ when $g$ is simple. Knowing all the
automorphism invariants for the unextended algebra is a necessary and major
step towards the full classification.

\bigskip Write $\r=r+1$ and
let $n$ be the height $k+\r$. We will usually identify a weight $\la$ with
its Dynkin labels $(\la_0;\la_1,\ldots,\la_r)$. E.g.\ $\rho=(k+1;1,\ldots
,1)$. Then $\la\in P_{++}(\A_{r,k})$ iff each $\la_i$ is a positive integer
and $\sum_{i=0}^{r}\la_i=n$. The outer automorphisms of $\A_{r,k}$ are
generated by charge conjugation $C$ (order 2, unless $r=1$ when $C=id.$)
and $A$ (order $\r$), defined by $C\la=(\la_0;\la_{r},\la_{r-1},\ldots,\la_2
,\la_1)$ and $A\la=(\la_{r};\la_0,\la_1,\ldots,\la_{r-1})$. Throughout
this paper, $S$ will denote the Kac-Peterson modular matrix [16] for
$\A_{r,k}$, corresponding to the transformation $\tau\rightarrow -1/\tau$.
First we write down some familiar equations:
$$\eqalignno{t(\la)\Eqde&\sum_{a=1}^{r} a\la_a,&(3a)\cr
t(A^a\la)\equiv &\, na+t(\la)\ ({\rm mod}\ \r),&(3b)\cr
(\overline{A^a\la})^2\equiv& \,\overline{\la}^2-2na\,t(\la)/\r+n^2a\,(\r-a)/
\r,\ ({\rm mod}\ 2n),&(3c)\cr
S_{A^a\la,A^b\mu}=&\exp[2\pi i\bigl(b\,t(\la-\rho)+a\,t(\mu-\rho)+nab\bigr)
/\r]\,S_{\la\mu},&(3d)\cr
S_{C\la,\mu}=&S_{\la,C\mu}=S_{\la\mu}^*,\qquad (\overline{C\la})^2=
\overline{\la}^2.&(3e)\cr}$$
By $\overline{\la}^2$ we mean the usual norm of the horizontal projection
$\overline{\la}=(\la_1,\ldots,\la_r)$.
Note that $t(\rho)\equiv 0$ (mod $\r$) for $r$ even, and $t(\rho)\equiv
\r/2$ (mod $\r$) for $r$ odd.

We are trying to find all permutations $\sig$ of $P_{++}(\A_{r,k})$ such that
$$\eqalignno{S_{\la\mu}=&S_{\sig\la,\sig\mu},\qquad \forall \la,\mu\in P_{++}
(\A_{r,k});&(4a)\cr
\overline{\la}^2\equiv &(\overline{\sig \la})^2\ ({\rm mod}\ 2n),\qquad \forall
\la\in P_{++}(\A_{r,k}).&(4b)\cr}$$
By abuse of language, any such $\sig$ will be called an {\it automorphism
invariant}. The simplest examples are $\sig=id.$ and $\sig=C$. It is important
to realize that all automorphism invariants will form a group under
composition. The relation between $\sig$ and the
physical invariant $M$ in eq.(1) is given by $M_{\la\mu}=\delta_{\mu,\sig
\la}$. From this and $M_{\rho\rho}=1$ we get that $\sig\rho=\rho$.

In this paper the characters $\chi$ in eq.(1) are taken with their full
$z$-dependence, so e.g.\ $\chi_\la=\chi_{C\la}$ only if $\la=C\la$. This
is in fact demanded by the mathematics: many serious mathematical
complications arise when, as sometimes appears in the literature, $z$
is set equal to 0.

Next let us write down the ``obvious'' automorphism
invariants of $\A_{r,k}$ [6]. Define
$$\tilde{n}=\left\{\matrix{n&{\rm if}\ r\equiv  n\equiv 0\ {\rm (mod\ 2)};
\cr k&{\rm otherwise}.\cr}\right. \eqno(5a)$$
Choose any positive integer $m$ dividing $\r$ such that
$m\tilde{n}$ is even and $gcd(\r/m,m\tilde{n}/2)=1$. Then  we
can find an integer $s$ such that $sm\tilde{n}/2\equiv 1$ (mod $\r/m$).
To each such divisor $m$ of $\r$, we can define an automorphism invariant
$\sig_m$ given by
$$\sig_m(\la)\Eqde A^{-s\,m\,t(\la-\rho)}\la.\eqno(5b)$$
For example, $\sig_{\r}$ corresponds to the identity permutation.
The conjugation $C\sig_m$ of any of these is also an automorphism invariant.
That each $\sig_m$ is a bijection, follows from $\sig_m\circ \sig_m=id.$.
That they satisfy eqs.(4) can be verified explicitly, using eqs.(3).

\medskip\noindent{\bf Theorem}:\quad The only automorphism invariants $\sig$
of $\A_{r,k}$ are $C^a\sig_m$, for $a=0,1$ and $\sig_m$ defined in eq.(5b).
All of these are distinct, except for $r=1$ (when $C=id.$), or when $k\le 2$.
\medskip

{}From the theorem we see that all automorphism invariants are of order 2
(i.e.\ $\sig^2=id.$), and commute: in fact
$$C^a\sig_m\circ C^b\sig_{m'}=C^{a+b}\sig_{m''},\ {\rm where}\ m''=\r \,
gcd(m,m')^2/mm'.\eqno(5c)$$
Both of these facts are surprising, and not true for general (semi-simple)
$g$. Three special cases of the theorem were known
previously: $r=1$ [1,10], $k=1$ [11], and $r=2$ [12,17]. We now proceed
to the proof of our theorem.

For each $i=1,\ldots,r$ define $\om^i=(k;1,\ldots,2,\ldots,1)$ -- i.e.\
$(\om^i)_j=1+\delta_{ij}$ for $0<j\le r$. This is the weight of the $i$th
fundamental representation. For any $\la$
define the outer automorphism orbit $[\la]=\{C^aA^b\la\}$. Let $o(\la)$
equal the number of indices $0\le i\le r$ such that $\la_i>1$. E.g.\
$\la\in[\rho]$ iff $o(\la)=1$. Also, $o(\om^i)=2$ (provided $k>1$).

Let's begin by proving the second statement in the Theorem.
That the $\sig_m$'s are all distinct is easy to see from $\sig_m(\om^1)$.
Finally, for $r>1$ and $k>2$, $C\om^1\not\in\{A^i\om^1\}$, so $C\sig_m(\om^1)
\ne \sig_{m'}$ for any $m,m'$. Incidently, when $k\le 2$, $C=\sig_m$ where
$m=1$ or 2. The number of automorphism invariants is counted in eq.(2)
above.

\medskip \noindent{{\bf Claim 1:}}\quad $\sig(\om^1)\in[\om^1]$.

\noindent{{\it Proof}}\quad The idea here is to show that although
$S_{\rho\la}$ is smallest for $\la\in[\rho]$, it is second smallest for
$\la\in[\om^1]$. From eq.(4a), this would prove Claim 1.

We have the following formula:
$$Q(\la)\Eqde {S_{\rho\la}\over S_{\rho\rho}}=\prod_{\bar{\alpha}>0}
{\sin[\pi\overline{\la}\cdot\overline{\alpha}/n]\over \sin[
\pi\overline{\rho}\cdot\overline{\al}/n]}.\eqno(6a)$$
The $\overline{\alpha}>0$ in eq.(6a) are the positive roots, which in the
case of $A_r$ look like $\sum_{j=\ell}^m
\overline{\alpha}_j$, where $\overline{\alpha}_j$ is the $j$th simple root,
and $1\le \ell\le m\le r$. $Q(\la)$ is constant within each orbit $[\la]$.
This expression was analysed in detail in [18]. Among other things, it
was found that $Q(\la)$, treated as a function of $\r$ {\it real} variables
$\la_i$, is concave-down (provided
the usual conditions $\la_i\ge 1$, $\sum_i \la_i=n$ are imposed).

So consider any $\la\in P_{++}(\A_{r,k})$, $\la\not\in[\rho]$. Suppose first
that
$o(\la)\ge 3$ -- say $\la_i>1$ and $\la_j>1$. Consider $\mu(t)=(\la_0;\la_1,
\ldots,\la_i-t,\ldots,\la_j+t,\ldots,\la_{r})$, so that $\mu(0)=\la$. Then
$(\mu(\la_i-1))_i=(\mu(1-\la_j))_j=1$, and for $1-\la_j\le
t\le \la_i-1$ $\mu(t)$ obeys the proper inequalities. An easy calculation,
similar to that in [18], tells us
$${{\rm d}\over {\rm d}t}Q(\mu(t))=0\,\Rightarrow\,{{\rm d}^2\over {\rm d}t^2}
Q(\mu(t))=-Q(\mu(t))\,{\pi^2\over n^2}\,\sum_{\overline{\alpha}>0}{(
\overline{a}\cdot\overline{
\alpha})^2\over \sin^2[\pi \overline{\rho}\cdot \overline{\alpha}/n]}<0,
\eqno(6b)$$
where $\overline{a}=\overline{\om^i}-\overline{\om^j}$. Thus
$Q(\mu(t))$ will take its minimum on the endpoints, i.e.\ either
$Q(\mu(\la_i-1))<Q(\la)$ or $Q(\mu(1-\la_j))<Q(\la)$. In either case,
we have found a $\mu\in P_{++}(\A_{r,k})$ with $Q(\mu)<Q(\la)$, and which has
$o(\mu)=o(\la)-1$.

Continuing inductively, we see it suffices to consider the weights with
$o(\la)=2$. The same argument allows us to put one of those two Dynkin
labels equal to 2. In other words, starting with any weight $\la\not\in
[\rho]$, we can find a $\om^i$ such that $Q(\la)\ge Q(\om^i)$, with
equality iff $\la\in[\om^i]$. All that remains is to compare $Q(\om^i)$
with $Q(\om^1)$.

We find from eq.(6a) that for all $1<\ell\le r$,
$${Q(\om^{\ell})\over Q(\om^{\ell-1})}=\prod_{i=1}^{\ell-1}{\sin[\pi{\ell-i
\over n}]\over\sin[\pi{\ell+1-i\over n}]}\,\prod_{j=\ell}^{r}{\sin[\pi{j+2
-\ell\over n}]\over\sin[\pi{j+1-\ell\over n}]}={\sin[\pi{r+2-\ell\over n}]
\over\sin[\pi{\ell\over n}]}.\eqno(6c) $$
Hence $Q(\om^1)<Q(\om^i)$ unless $i=1,r$. But $\om^{r}=C\om^1$.\qquad  QED
\medskip

So $\sig(\om^1)=C^aA^b\om^1$ for some $a,b$. By hitting $\sig$ with $C^a$,
we may assume $a=0$. Evaluating the expression $(\overline{\om^1})^2\equiv (
\overline{A^b\om^1})^2$ (mod $2n$) using eq.(3c), we get
$$2b(1+t(\rho))\equiv nb(\r-b)\ ({\rm mod}\ 2\r).\eqno(7)$$

\medskip\noindent{{\bf Claim 2:}}\quad Let $b$ be any integer satisfying
eq.(7). Then $\sig_m(A^b\om^1)=\om^1$, for one of the $\sig_m$ of eq.(5b).

\noindent{{\it Proof}}\quad Given any solution $b$ to eq.(7), define $m=gcd(
b,\r)$ and $s'=b/m$ -- we can suppose (adding
$\r$ to $b$ if necessary) that $s'$ will be coprime to $2\r/m$.
We want to show two things: (i) $\tilde{n}m/2$ is an integer coprime
to $\r/m$ (so that $\sig_m$ exists); and (ii) $b\equiv -sm$ (mod $\r$) (so
that $\sig_m(A^b\om^1)=\om^1$). (ii) is equivalent to showing $\tilde{n}b/2
\equiv -1$ (mod $\r/m$), which implies (i).

Consider first $\r$ odd. Then dividing eq.(7) by $m$ gives $2s'+ns'b\equiv 0$
(mod $2\r/m$). This reduces to $nb\equiv -2$ (mod $2\r/m$), which is
equivalent to (ii).

For $\r$ even, we get $2+nb\equiv \r(1+n)$ (mod $2\r/m$). Thus it suffices
to show that $\r b\equiv \r(1+n)$ (mod $2\r/m$), i.e.\ that either $m$  or
$1+n+b$ is even. But $m$ odd means $b$ is odd, which by eq.(7) requires
$n$ even. \qquad QED  \medskip

By Claim 2 we see that, replacing $\sig$ with $\sig_m\circ\sig$, {\it it
suffices to consider those $\sig$ fixing
$\om^1$}. Our theorem is proved if we can show such a $\sig$ must equal the
identity.

By Verlinde's formula [19] and eq.(4a), $\sig$ must be
a symmetry of the fusion rules: $N_{\la\mu}^\nu=N_{\sig\la,\sig\mu}^{\sig
\nu}$. So it is natural to look at the fusions involving $\om^1$:
$$N_{\la\om^1}^\nu=\left\{\matrix{1&{\rm if}\ \nu\in\la+\{\om^1-\om^0,
\om^2-\om^1,\ldots,\om^0-\om^{r}\}\cr 0&{\rm otherwise}\cr}
\right.,\eqno(8a)$$
where $\om^0=\rho$. Note that
$$\sum_{\nu\in P_{++}(\A_{r,k})}N_{\la\om^1}^\nu=o(\la).\eqno(8b)$$
Because $\sig\om^1=\om^1$, and $\sig$ is a
permutation of $P_{++}(\A_{r,k})$, we get from eqs.(8a),(8b) respectively that
$$\eqalignno{N_{\sig\la,\om^1}^{\sig\nu}=&N_{\la,\om^1}^\nu,&(8c)\cr
o(\la)=&o(\sig\la).&(8d)\cr}$$

We are now ready to show $\sig$ must fix all other $\om^i$. $N_{\om^1,
\om^1}^\nu=1$ iff either $\nu=\om^2$ or $\nu=\om^{1,3}\Eqde
(k-1,3,1,\ldots,
1)$. This means, from eq.(8c), that $\sig\om^2$ must equal one of those two
weights. But the possibility $\sig\om^2=\om^{1,3}$ is eliminated by
eq.(4b): $(\overline{\om^2})^2=2(r+2)(r-1)/\r+r\r(r+2)/12$ but $(
\overline{\om^{1,3}})^2 =2r(r+3)/\r+r\r(r+2)/12$. Therefore $\sig\om^2=\om^2$.

Now suppose for some $1<i<r$ we know that $\sig\om^i=\om^i$. Then look
at $N_{\om^i,\om^1}^\nu$: it equals 1 iff either $\nu=\om^{i+1}$ or
$\nu=(k-1,2,1,\ldots,2,\ldots,1)$. These have $o(\nu)=2,3$ respectively.
So $\sig\om^{i+1}=\om^{i+1}$ is forced from eqs.(8c),(8d).

Thus $\sig$ must fix $\om^1,\ldots,\om^r$. From this
we can now complete the proof of the theorem, with the following
observation:

\medskip\noindent{\bf Claim 3}:\quad Suppose we have weights $\la,\la'\in
P_{++}(\A_{r,k})$ such that $S_{\om^i\la}/S_{\rho\la}=S_{\om^i\la'}/
S_{\rho\la'}$ for all $i$. Then $\la=\la'$.

\noindent{\it Proof}\quad We know [18] that each fusion matrix $N_\al$ can be
written as a polynomial $p_\al$ of $N_{\om^1},\ldots,N_{\om^{r}}$. We
also know [19] all $N_\al$ are simultaneously diagonalized by $S$: $S^{-1}
N_\al S=D_\al$, with eigenvalues $(D_\al)_{\be\be}=S_{\al\be}/S_{\rho
\be}$. Thus
$${S_{\al\be}\over S_{\rho\be}}=p_\al({S_{\om^1\be}\over S_{\rho\be}},
\ldots,{S_{\om^{r}\be}\over S_{\rho\be}}).\eqno(9a)$$
Together with the hypotheses of Claim 3, eq.(9a) tells us that in fact
$S_{\al\la}/S_{\rho\la}=S_{\al\la'}/S_{\rho\la'}$ for all $\al$. Hence
$$N_{\la\mu}^\nu=\sum_{\al\in P_{++}(\A_{r,k})}{S_{\la\al}\,S_{\mu\al}\,
S_{\nu\al}^*
\over S_{\rho\al}}={S_{\rho\la}\over S_{\rho\la'}}\sum_{\al\in P_{++}
(\A_{r,k})}{S_{\la'\al}\,S_{\mu\al}\,S_{\nu\al}^*\over S_{\rho\al}}=
{S_{\rho\la}\over S_{\rho\la'}}N_{\la'\mu}^\nu.\eqno(9b)$$
Putting $\mu=\rho$, $\nu=\la$ in eq.(9b), we get $1=\delta_{\la\la'}\,
S_{\rho\la}/S_{\rho\la'}$, which forces $\la=\la'$. \qquad QED\medskip

Choose any $\la\in P_{++}(\A_{r,k})$ and define $\la'=\sig\la$. Then from
eq.(4a) and
using $\sig\rho=\rho$ and $\sig\om^i=\om^i$ we find that the conditions of
Claim 3
are satisfied. Then Claim 3 tells us $\la=\la'$. In other words, $\sig$
must be the identity, and we are done!

\bigskip
\centerline{\bf Appendix: The fusion ring of $\widehat{SU}(m)_k$}\medskip

The conclusion to the proof brings to mind the fusion ring of $\A_{r,k}$
[20], and in particular the possibility of representing it with polynomials
in only one variable [21].

Call the pair $(r,k)$ ``non-degenerate'' if the eigenvalues of $N_{\om^1}$,
namely
$$D^1_{r,k}(\la)={S_{\om^1\la}\over S_{\rho\la}}=\zeta^{t(\la)}
\sum_{m=0}^{r}\zeta^{\r\sum_{j=0}^{m}\la_j},\qquad \zeta=\exp[2\pi i/\r
n],\eqno(10a)$$
$\forall \la\in P_{++}(\A_{r,k})$, are all distinct. As before, $\r=r+1$ and
$n=\r+k$. Then $(r,k)$ will be
non-degenerate iff the fusion ring of $\A_{r,k}$ can be generated by
polynomials in the fusion matrix $N_{\om^1}$ [21]. It is easy to show
that $(r,1)$ and $(1,k)$ will always be non-degenerate. [21] showed that
$(2,k)$ is as well -- in fact this follows trivially from our Claim 3.
However, not all pairs $(r,k)$ are non-degenerate: [21] gave (3,2) as
an example.

It should be possible to completely solve this problem, i.e.\ to find all
non-degenerate $(r,k)$. In this appendix we collect together a few results
in this direction, which give several examples of degenerate and
non-degenerate pairs $(r,k)$.\medskip

\item{{\bf (a)}} Suppose $gcd(\r,k)>1$ and $\r$ is not prime. Then
$(r,k)$ will be degenerate -- in particular more than one $\la$ will have
$D^1_{r,k}(\la)=0$.

\item{{\bf (b)}} Choose any prime $p$ dividing $n$ ($p$ may or may not
divide $\r$). Then $(r,k)$ will be degenerate if $2p\le \r\le n-2p$.

\item{{\bf (c)}} Suppose $n=p^\ell$ for some prime $p$. Then $(r,k)$ will
 be degenerate iff either $2p\le \r\le p^\ell-2p$  or $k=p\ne \r$.

\item{{\bf (d)}} Suppose $(r,k)$ is non-degenerate. Then $(k-1,r+1)$
will be non-degenerate iff either $gcd(\r,k)=1$ or $\r=k$.\medskip

(a) and (b) tell us that for $r>2$, then there will be
infinitely many $k$ such that $(r,k)$ is degenerate. (c) tells us that
$(r,k)$ will be non-degenerate whenever $n$ is prime.
(d) tells us that $(r,2)$ will be non-degenerate iff either $r$ is even or
$r=1$, and that $(r,3)$ will be non-degenerate iff either $r\equiv 0,1$
(mod 3) or $r=2$.

(a) is proved by construction: let $p$ be a prime dividing $k$ and $\r$.
Choose any $\r/p-1$ distinct integers $\ell_i$ between 0 and $n/p$
exclusive. Put $\ell_0=0$, and for $1\le j\le p-1$ and $0\le i<\r/p$ define
$\ell_{i+j\r/p}=\ell_i+jn/p$. For each $1\le i\le r$ write $\la_i=\ell_i-
\ell_{i-1}$. Then $D^1_{r,k}(\la)=0$.

The construction for (b) is similar, except that here $\zeta^{-t(\la)-\r
\la_0} D^1_{r,k}(\la)-\zeta^{-t(\la)-\r\la_0}D^1_{r,k}(\mu)=0$ must be
constructed, instead of merely $\zeta^{-t(\la)-\r\la_0}D^1_{r,k}(\la)=0$
as was done for (a).

The non-degeneracy in (c) is proved by noting the following immediate
consequence of eq.(10a): write all exponents $e_j$ of $\zeta$ in $D^1_{r,k}
(\la)=D^1_{r,k}(\mu)$ so that $0\le e_i<\r n$; if we then replace $\zeta$
in $D^1_{r,k}(\la)=D^1_{r,k}(\mu)$ with the variable $x$, and the resulting
polynomial equation is {\it identically} satisfied, then $\mu=A^i\la$ for
some $i$. If in addition $D^1_{r,k}(\la)\ne 0$, then $\la=\mu$.

(d) is a consequence of rank-level duality [22]. In particular we get that
$$D^1_{r,k}(\la)=\exp[2\pi i\{t(\la-\rho)+\r\,(\la_0-1)\}/\r k]\,
D^1_{k-1,r+1}(C\beta\la).\eqno(10b)$$
where $\beta$ is the map $P_{++}(\A_{r,k})\rightarrow P_{++}(\A_{k-1,r+1})$
defined in [22].

\medskip
These results tell us that $SU(2)$ and $SU(3)$ are the only $SU(m)$ whose
fusion ring {\it at all levels} $k$ can be represented by polynomials in
only one variable $x$. For each $m>3$, there will be infinitely many $k$
for which the fusion ring of $\widehat{SU}(m)_k$ requires more than one
variable $x_i$, and infinitely
many other $k$ for which one variable will suffice.

\bigskip\noindent{{\bf Acknowledgements.}} I am especially grateful to
Philippe Ruelle
for convincing me that this problem is solvable, and for reminding me
of [18], and to Mark Walton for his patient explanations of fusion rules.
Also, Antoine Coste and Jean-Bernard Zuber were helpful with general
information. The hospitality of IHES is also appreciated.\bigskip

\item{[1]} A.\ Cappelli, C.\ Itzykson, J.-B.\ Zuber, {\it Commun.\ Math.\
Phys.}\ {\bf 113} 1-26 (1987)

\item{[2]} Ph.\ Ruelle, E.\ Thiran, J.\ Weyers, {\it Nucl.\ Phys.}\
{\bf B402} 693-708 (1993)

\item{[3]} J.-B.\ Zuber, talk given at ICMP, Paris, 1994

\item{[4]} D.\ Altschuler, J.\  Lacki, Ph.\ Zaugg, Ph., {\it Phys.\ Lett.}\
{\bf B205} 281-284 (1988)

\item{[5]} D.\ Bernard, {\it Nucl.\ Phys.}\ {\bf B288} 628-648 (1987)

\item{[6]} A.\ N.\ Schellekens, S.\ Yankielowicz, {\it Nucl.\ Phys.}\ {\bf
B327} 673-703 (1989); {\it Phys.\ Lett.}\ {\bf B227} 387-391 (1989)

\item{[7]} P.\ Bouwknegt, W.\ Nahm, {\it Phys.\ Lett.}\ {\bf B184} 359-362
(1987);

\item{} F.\ A.\ Bais, P.\ G.\ Bouwknegt, {\it Nucl.\ Phys.}\ {\bf B279}
561-570 (1987);

\item{} A.\ N.\ Schellekens, N.\ P.\ Warner, {\it Phys.\ Rev.}\ {\bf D34}
3092-3096 (1986)

\item{[8]} M.\ A.\ Walton, {\it Nucl.\ Phys.}\ {\bf B322} 775 (1989);

\item{} D.\ Altschuler, M.\ Bauer, C.\ Itzykson, {\it Commun.\ Math.\ Phys.}\
{\bf 132} 349 (1990);

\item{} D.\ Verstegen, {\it Nucl.\ Phys.}\ {\bf B346} 349 (1990);

\item{} A.\ Font, {\it Mod.\ Phys.\ Lett.}\ {\bf A6} 3265 (1991);

\item{} M.\ R.\ Abolhassani, F.\ Ardalan, ``A unified scheme for modular
invariant partition functions of WZW models.'' (May 1993)

\item{[9]} M.\ Kreuzer, A.\ N.\ Schellekens, {\it Nucl.\ Phys.}\ {\bf B411}
97-121 (1994)

\item{[10]} A.\ Cappelli, C.\ Itzykson, J.-B.\ Zuber, {\it Nucl.\ Phys.}\
{\bf B280} 445-465 (1987);

\item{} D.\ Gepner, Z.\ Qui, {\it Nucl.\ Phys.}\ {\bf B285} 423 (1987);

\item{} A.\ Kato, {\it Mod.\ Phys.\ Lett.}\ {\bf A2} 585 (1987)

\item{[11]} C.\ Itzykson, {\it Nucl.\ Phys.\ (Proc.\ Suppl.)} {\bf 5B}
150-165 (1988);

\item{} P.\ Degiovanni, {\it Commun.\ Math.\ Phys.}\ {\bf 127} 71-99 (1990)

\item{[12]} T.\ Gannon, {\it Commun.\ Math.\ Phys.}\ {\bf 161} 233-264 (1994);
``The classification of SU(3) modular invariants revisited.'' IHES preprint
P/94/32 (hep-th/9404185)

\item{[13]} T.\ Gannon, Ph.\ Ruelle (work in progress)

\item{[14]} T.\ Gannon, ``Towards a classification of SU(2)$\oplus\cdots
\oplus$SU(2) modular invariant partition functions. IHES preprint P/94/21
(hep-th/9402074)

\item{[15]} B.\ Gato-Rivera, A.\ N.\ Schellekens, {\it Nucl.\ Phys.}\
{\bf B353} 519-537 (1991)

\item{[16]} see e.g.\ V.\ G.\ Kac, {\it Infinite Dimensional Lie Algebras},
3rd ed. (Cambridge University Press, 1990)

\item{[17]} Ph.\ Ruelle, {\it Commun.\ Math.\ Phys.}\ {\bf 160} 475-492
(1994)

\item{[18]} J. Fuchs, {\it Commun.\ Math.\ Phys.}\ {\bf  136}, 345-356 (1991)

\item{[19]} E.\ Verlinde, {\it Nucl.\ Phys.}\ {\bf B[FS22] 300} 360-376
(1988);

G.\ Moore, N.\ Seiberg, {\it Phys.\ Lett.}\ {\bf B212} 451-460 (1988)

\item{[20]} see e.g.\ D.\ Gepner, {\it Commun.\ Math.\ Phys.}\ {\bf 141}
381 (1991)

\item{[21]} P.\ Di Francesco, J.-B.\ Zuber, {\it J.\ Phys.}\ {\bf A26}
1441-1454 (1993)

\item{[22]} D.\ Altschuler, M.\ Bauer, C.\ Itzykson, {\it Commun.\ Math.\
Phys.}\ {\bf 127} 349-364 (1990)

\end